\documentclass{article}          %% The manuscript based on AASTeX v5.x
 \usepackage{graphicx}
\usepackage{hyperref}
\usepackage[margin=2.5cm]{geometry}

\begin{document}

%% Article title
%
\title{\bf Cross correlation and time-lag between cosmic ray intensity and solar activity during solar cycles 21, 22 and 23}

%% Running heads
\author{D. Sierra-Porta \\ \small Grupo de Investigaci\'on en Relatividad y Gravitaci\'on (GIRG), \\ \small Escuela de F\'{\i}sica, Universidad Industrial de Santander, 680002 Bucaramanga-Colombia and \\ \small Centro de Modelado Cient\'ifico (CMC), Universidad del Zulia, 4001 Maracaibo-Venezuela. \\ Corresponfing author: dsierrap@uis.edu.co} %% non-output

%\altaffiltext{2}{}
%\altaffiltext{3}{}
\maketitle

%% Abstract
\begin{abstract}
In the present paper a systematic study is carried out to validate the similarity or degree of relationship between daily terrestrial cosmic rays intensity and three characteristics of about evolution of solar corona, like a number of sunspots and flare index observed in the solar corona and to Ap index for regular magnetic field variation caused by regular solar radiation changes. The study is made in a range including three solar cycles starting with the cycle 21 (year 1976) and ending on cycle 23 (year 2008). The technique used in this case will be the use of the cross-correlation technique to establish patterns and dependence on the behavior of both variables. This study focused on the time lag calculation for these variables and found a maximum of negative correlation over $CC_1\approx 0.85$, $CC_2\approx 0.75$ and $CC_3\approx 0.63$ with an estimation of 181, 156 and 2 days of deviation between maximum/minimum of peaks for the intensity of cosmic rays related with sunspot number, flare index and Ap index regression, respectively.
\\
\textbf{keywords:} Space weather -- Intensity of Cosmic Rays -- Solar activity.

\end{abstract}

%% Keywords

\section{Aims and scope}
The following is a study of Cross-Correlation between Cosmic Ray Intensity (ICR) and a few characteristics that determine the evolution on solar activity taking into consideration total daily Sunspot Number (SN), Flare Index (FI) on the solar corona and Ap-Index (ApI) for the regular magnetic field variation caused by regular solar radiation changes. Although SN is not the only one of the parameters that determine the dynamics of the solar coronal and therefore the activity of the sun, these are used as one of the reliable and easily available solar parameters that provide information about the variability of solar activity and they are convenient and represent a good measurement parameter that can be combined and complemented with others for a more accurate understanding as a solar parameter in cosmic ray studies. The numbers of sunspots as an active parameter have been used for research mainly because it is known that solar flares emanate from the places where sunspots occur and eventually scatter in the direction of the earth. Galactic cosmic rays in the energy range from several hundred MeV to a few GeV are subject to heliospheric modulation because solar production and its variation affect its intensity and spectrum during the solar activity cycle of about 11 years \cite{schwabe1844sunspot}. It is well known that the variation of intensity of cosmic rays shows an inverse correlation with the number of sunspots. But, in general, it is observed that the maximum/minimum number of sunspots does not coincide with the minimum/maximum intensity of the cosmic ray. 

Long-term cosmic ray observations by means of neutron monitors now allow us to distinguish modulation effects related to the 22-year solar magnetic cycle from effects due to the 11-year solar activity cycle. Popielawska and others \cite{popielawska1995cosmic} \cite{popielawska1992components} have reported a positive correlation when studying the variations of solar cycles of around 11 years and cycles of magnetic activity of around 22 years using observations of a pair of Climax\footnote{Latitude 39.37$^{\circ}$ N, longitude 106.18$^{\circ}$ W, elevation 1612 m, rigidity (1965) 2.99 GV, standard atmospheric pressure: 672 mb.}-Huancayo\footnote{Latitude 12.0$^{\circ}$ S, longitude 75.3$^{\circ}$ W, elevation 3350 m, rigidity (1965) 12.92 GV, standard atmospheric pressure: 704 mb.} and confirmed for Kiel\footnote{Latitude	54.30$^{\circ}$ N, longitude 10.10$^{\circ}$ E, elevation 54 m, rigidity (1965) 2.36 GV, standard atmospheric pressure: 1006.6 mb.}-Tsumeb\footnote{Latitude 19.20$^{\circ}$ S, longitude 17.58$^{\circ}$ E, elevation 1240 m, rigidity (1965) 9.21 GV, standard atmospheric pressure: 880.0 mb.} neutron detectors with different cutoff rigidity, latitudes and longitudes, observing a relation of dependence between the variations of solar cycles and magnetic cycles associated with polarity changes of the latter. The final goal was to get an idea about the modulation of cosmic rays and other fundamental characteristics of cosmic rays.

In another report \cite{mishra2005solar} has also made a detailed study in this case by measuring the correlation between different types of observables for the modulation of solar cycles and the intensity of cosmic rays. For this case they have used large datasets with monthly resolution between years 1950 and 2003 corresponding to solar cycles 20 to 23. In this report, the pressure corrected monthly values of cosmic rays are obtained from the data of Kiel neutron monitors for the period 1950-2003. The mean values of sunspot number have been taken from the solar geophysical data. They observe a negative and high correlation between solar activity and cosmic rays. The interest in studying correlations of cosmic rays with different types of indicators of solar activity is interesting from the point of view of understanding the interrelationships between both variables \cite{dorman1967solar} \cite{usoskin1998correlative} \cite{van2000modulation}.

Many other researchers have studied this phenomenon taking into account the hysteresis effect between these two variables and other parameters of the solar activity as well that is useful for the estimation of the modulating region of the heliosphere using in this case monthly values for indices and cosmic ray intensity, for a revised of this topic we can forward to \cite{mavromichalaki1984time}, \cite{mavromichalaki1998simulated}, \cite{mavromichalaki2007cosmic} and \cite{paouris2012galactic} and references therein. An interesting particularity of these works includes the generation of new indices constructed as a linear combination of individual indices using for such a calculation of the maximum correlation for the intensity of cosmic rays, using neutron monitor detectors, a distinction between even and odd solar cycles, as well as between the declining and ascending phases of them, is well established by this authors.

Recently, a new statistical technique called "cross-correlation" widely used in many fields and applications has been achieved to study the correlation between SN and ICR. The main motivation of this manuscript is to start the study, with the application of this technique to try to study the correlation between ICR and solar activity parameters and using them in the statistical technique for different periods or solar cycles. This study provide a calculation for recent cycles of sun activity and using data of ICR in above equatorial region. In this works additionally to this index we introduce two more index for a more complete study of these correlation including flare index observed in the solar corona and to Ap index for regular magnetic field variation caused by regular solar radiation changes.

\section{Data, methods, and results}
To carry out the task, we have compiled a dataset of existing and open source data: Cosmic Ray Intensity Data (daily) available from 1957 to the present provided by the Kiel Neutron Monitor, total daily Number of Sunspots (daily) available between the 1818 to current years from WDC-SILSO, Royal Observatory of Belgium, Brussels (\url{http://www.sidc.be/silso/datafiles}), Interplanetary Ap index provided from Helmholtz Centre Potsdam (GFZ German Research Centre For Geosciences) and Flare Index Data used in this study were calculated by T.Atac and
A.Ozguc from Bogazici University Kandilli Observatory, Istanbul, Turkey. We have selected between the same range of available years for cosmic ray intensity to do the corresponding validation and correlation.

The data allows a study of the correlation for the solar cycle 21 (from March 1976, duration 10.3 yr), cycle 22 (from August 1986, duration 10.0 yr), and cycle 23 (from September 1996, duration 12.2 yr). The resolution of the data available is daily. For the case of ICR data, the number of daily total counts/min not corrected by pressure and a set of data of the same type corrected by pressure are available. As usual when working with experimental data, a process of regularization, cleaning and normalization of data is carried out in order to eliminate erroneous, missing or critical data points.

\begin{figure}[!ht]
\centering
\includegraphics[width=0.6\columnwidth]{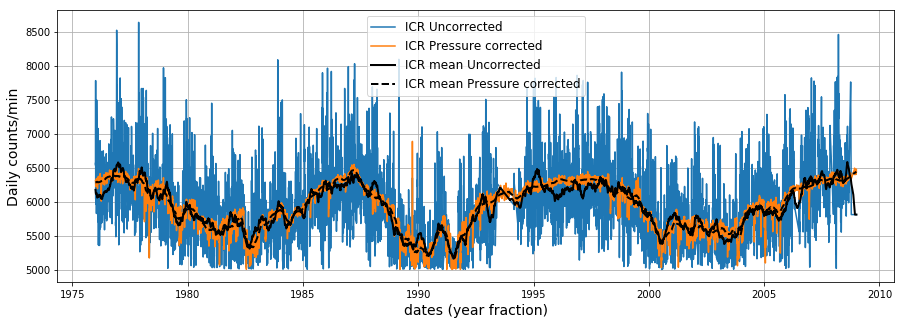}
\includegraphics[width=0.6\columnwidth]{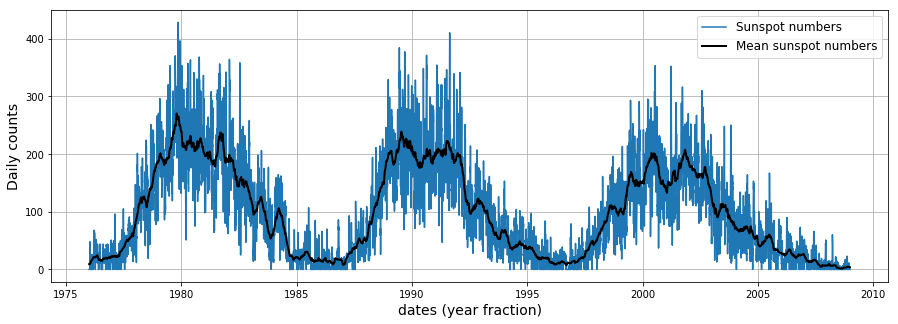}
\includegraphics[width=0.6\columnwidth]{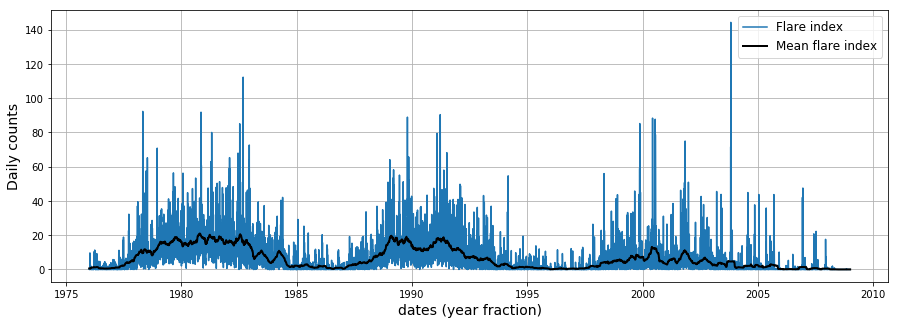}
\includegraphics[width=0.6\columnwidth]{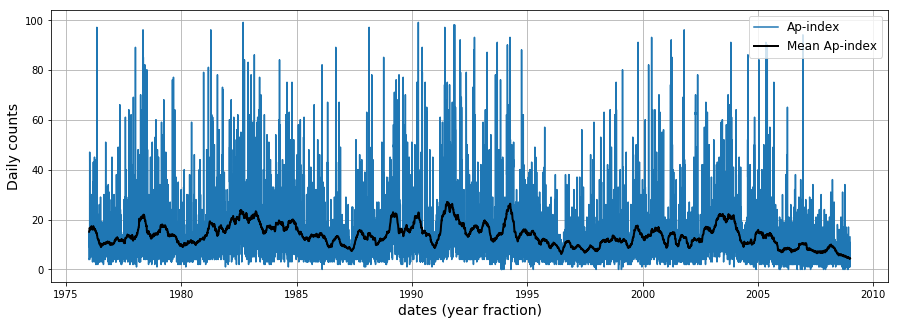}
\caption{Intensity of cosmic rays for the period of years between 1976 and 2008. The horizontal axis is found in units of fractions of years. The vertical column represents the actual value of counts/min for cosmic ray intensity.}
\label{IRC_plot}
\end{figure}

To measure the relationship between cosmic rays and these solar and geomagnetic activity indices for the solar cycle (21 to 23), the correlation coefficient between the daily mean values of these two variables is calculated. Figures \ref{IRC_plot} show the variations of all parameters during the period considered.

One thing that could be done is make data normalizing for placing the two variables in the same units and orders of magnitude, however, although this process could be done simply by applying a Fourier transform on each series and then dividing each one by the value of the zero mode of the same, this is not necessary and does not influence either the quality or validity of the results.

The study carried out consists in the calculation of the cross correlation coefficient for the two series. Mathematically and technically the calculation allows to determine the degree of similarity and/or relationship between two variables as a function of the relative displacement between the two. This can be seen in essence as a convolution of a data set. Strictly speaking assuming two series $f[i]$ and $g[i]$ then the correlation is defined as
\begin{equation}
(F\star G)[i]:=\sum _{j=-\infty }^{\infty }F^{*}[j]G[i+j].
\end{equation}

Cross-correlations are useful for determining the time delay between two signals. The maximum (or minimum if the signals are negatively correlated) of the cross-correlation function indicates the point in time where the signals are best aligned. The time delay between the two signals is determined by the argument of the maximum, or {\textbf arg max} of the cross-correlation, as in
\begin{equation}
\tau_{\mathrm{delay}}=\mathrm{\mathbf{arg\,max}}\left[(F\star G)[i]\right]\Big|_i.
\end{equation}

\begin{figure}[!ht]
\centering
\includegraphics[width=0.6\columnwidth]{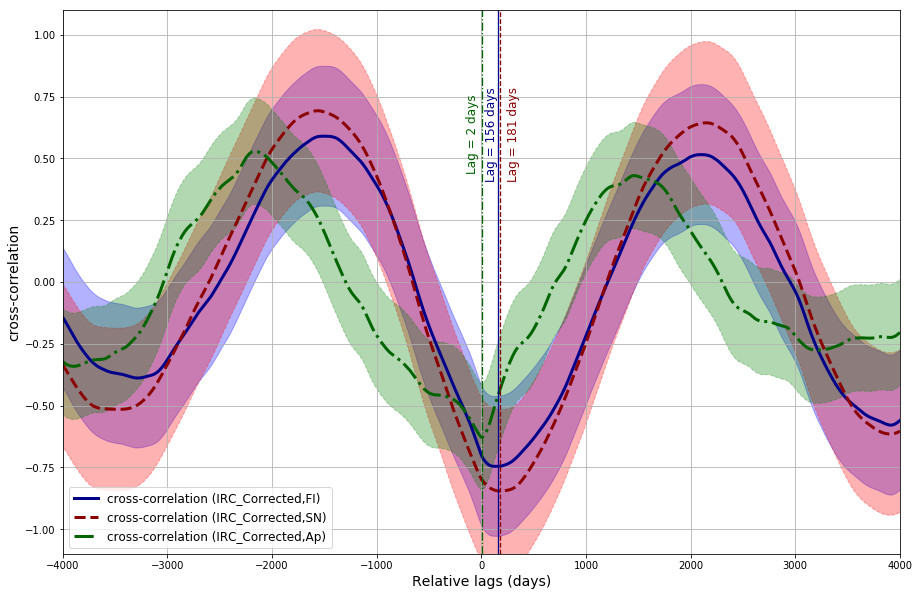}
\caption{Cross-Correlation function for four variables considered for the period of years between 1976 and 2008. The horizontal axis is found in units of symmetric fractions of years for the lag value.}
\label{CC_plot}
\end{figure}

The running cross-correlation function (CC) between pressure correction ICR and (SN, FI and Ap) for neutron observatory of Kield is shown in Figure \ref{CC_plot}. From the period 1976-2008 one can see a localized behavior of cross-correlation function. It is observed from Figure \ref{CC_plot} that the anti-correlation between IRC and SN, FI and Ap is strong ($>0.85$, $>0.75$ and $>0.63$) on a lag of 181, 156 and 2 days, respectively, while it becomes weak ($0-|0.25|$) during the rest of range. 

\section{Conclusion and discussion}
Solar activity rises and falls with a period of about 11 years. The number of sunspots indicates the level of solar activity. Emissions of matter and electromagnetic fields from the Sun increase during high solar activity, making it harder for Galactic cosmic rays to reach Earth. Solar activity is currently low, and the cosmic ray intensity is high. Soon the sunspots will increase and cosmic ray intensity will start to decrease.

This present work revealed that there is a strong negative relationship between Cosmic Rays Intensity and SunsPot Numbers taking into account detailed data sets from Kiel Neutron Monitor and monitors of solar activity (Brussels Royal Observatory), Flare index (Bogazici University Kandilli Observatory), including to Interplanetary Ap index provided from Helmholtz Centre Potsdam (GFZ German Research Centre For Geosciences). Also the negative relationship is revealed at lag 181 days, that is, maximum/minimum intensity of cosmic rays are preceded by minimum/maximum number of solar number spots for a time close to 181 days. As is known, this time lag is mainly due to a diffusive and dragging process of cosmic ray particles in the heliosphere which is considerably large under the influence of the solar wind and its magnetic field, which is composed of ions from the atmosphere solar \cite{dorman1967solar}. One of the values added in this paper is the use of data with daily resolution both for the number of sunspots and for the intensity of cosmic rays. 

The Cross-Correlation Function further revealed a stable relationship between maximum and minimum for both variables. Other variable in consideration reveal a strong correlation, in this case with 156 day on the time-lag respect to intensity of cosmic rays and Flare index. Both of these solar activity variables become a very good candidates for estimating cosmic rays modulation from solar indices. In other words, geomagnetic effects on the correlation taking into account Ap index is a good approximation in this study with a correlation over 0.65, nevertheless, it is possible refine this approach including other geomagnetic-like indices for a greater degree of correlation.

In conclusion we can say that in the period considered of three solar cycles (21-23) between 1976-2008, the solar indices as well as the index of geomagnetic activity in interstellar space have turned out to be good estimators of the cosmic ray modulation in this period. For the solar activity indices there are time-lags over 150 days (very close both) while for the modulation by geomagnetic activity a very small time-lag.

From this analysis, it is concluded that SN is a good index to choose for any long-term study on the variation of cosmic rays. Using data with a resolution of one day, it is possible to extend this study to consider other solar index that allow an analysis of which of them influence, to a greater or lesser extent, the variation of cosmic ray intensity. In this case, two solar cycles have been studied but the study could be extended to include a larger interval and another solar cycle. This study is ongoing at this time.

\section{acknowledgments}
Part of this work was supported by the Vicerrector\'{\i}a Investigaci\'on y Extensi\'on Universidad Industrial de Santander for its permanent sponsorship. DSP wants to thank the GIRG, Grupo Halley  and Vicerrector\'{\i}a Investigaci\'on y Extensi\'on of Universidad Industrial de Santander for the hospitality during my post-doctoral fellowship. DSP also thanks AGMS for the comments, discussions and help in applying the technique to large amounts of datasets. The author thanks to an anonymous reviewer for the suggestion of some references of some studies done previously, for the suitable comments and discussions.

\bibliographystyle{alpha}
\bibliography{sample}

%% References
%% Please cite all reference entries in the article text using \cite or
%% equivalent command. 

%%%  Using BibTeX  (Name-Year style)
%
% \bibliographystyle{spr-mp-nameyear-cnd}  %% BibTeX style
% \bibliography{<bib data>}                %% BibTeX data

%% Non-BibTeX  (Name-Year style)
%
% \begin{thebibliography}{}
% \bibitem[\protect\citeauthoryear{<author>}{<year>]{ref:?}
%    <ref. entry>
% \bibitem[\protect\citeauthoryear{<author>}{<year>]{ref:?}
%    <ref. entry>
% \end{thebibliography}

\end{document}